\documentstyle[12pt]{article}
\def\laq{\raise 0.4 ex \hbox{$<$}\kern -0.8 em\lower 0.62 ex\hbox{$\sim$}}
\def\gaq{\raise 0.4 ex \hbox{$>$}\kern -0.7 em\lower 0.62 ex\hbox{$\sim$}}
 
\def\beq{\begin{equation}}
\def\eeq{\end{equation}}
\def\bea{\begin{eqnarray}}
\def\eea{\end{eqnarray}}
\def\bq{\begin{quote}}
\def\eq{\end{quote}}

\def\al{\alpha}
\def\be{\beta}
\def\ga{\gamma}
\def\de{\delta}

\def\et{\eta}

\def\rh{\rho}

\def\si{\sigma}

\def\ph{\phi}

\def\ps{\psi}
\def\om{\omega}

\def\De{\Delta}

\def\prt{\partial}

\def\vev#1{\langle {#1}\rangle}

\def\half{{\textstyle{1\over 2}}}
\def\frac#1#2{{\textstyle{{#1}\over {#2}}}}

\def\lsim{\mathrel{\rlap{\lower4pt\hbox{\hskip1pt$\sim$}}
    \raise1pt\hbox{$<$}}}
\def\gsim{\mathrel{\rlap{\lower4pt\hbox{\hskip1pt$\sim$}}
    \raise1pt\hbox{$>$}}}
\def\sqr#1#2{{\vcenter{\vbox{\hrule height.#2pt
         \hbox{\vrule width.#2pt height#1pt \kern#1pt
         \vrule width.#2pt}
         \hrule height.#2pt}}}}

\def\rl{\stackrel{\leftrightarrow}{\hskip2pt\prt^{\mu}}}

\def\rln{\stackrel{\leftrightarrow}{\hskip2pt\prt^{\nu}}}
 
\def\AJ{{\it Ap. J.} }

\def\APP{{\it Acta Phys. Pol.} }

\def\APP{{\it Astropart. Phys.} }

\def\CQG{{\it Class. Quantum Gravity} }

\def\JP{{\it J. Phys.} }

\def\NAT{{\it Nature} }

\def\NP{{\it Nucl. Phys.} }
\def\PL{{\it Phys. Lett.} }
\def\PR{{\it Phys. Rev.} }
\def\PRL{{\it Phys. Rev. Lett.} }
\def\PRTS{{\it Phys. Rep.} }

\def\PMAG{{\it Philos. Mag.} }

\def\gappeq{\mathrel{\rlap {\raise.5ex\hbox{$>$}}
{\lower.5ex\hbox{$\sim$}}}}
 
\def\lappeq{\mathrel{\rlap{\raise.5ex\hbox{$<$}}
{\lower.5ex\hbox{$\sim$}}}}
 
\begin{document}
\pagestyle{empty}
 
\begin{flushright}
{\sc DF/IST-3.99} \\
{\sc October 1999} \\
{\tt gr-qc/yymmnn}
\end{flushright}

\vspace*{0.5cm}
 
\begin{center}
{\large\bf Proposed astrophysical test of Lorentz invariance}\\
\vspace*{1.0cm}
{\bf O. Bertolami}\footnote{E-mail address: {\tt orfeu@cosmos.ist.utl.pt}} and 
{\bf C.S. Carvalho}\\
\medskip
{Departamento de F\'\i sica}\\
{Instituto Superior T\'ecnico, Av. Rovisco Pais}\\
{1049 - 001 Lisboa, Portugal}\\

\vspace*{1.5cm}
{\bf ABSTRACT} \\ 
\end{center}
\indent
 
\setlength{\baselineskip}{0.6cm}
 
Working in the context of a Lorentz-violating extension of the standard 
model we show that estimates of Lorentz symmetry violation 
extracted from ultra-high energy cosmic rays beyond the 
Greisen-Kuzmin-Zatsepin (GZK) cutoff 
allow for setting bounds on parameters of that 
extension. Furthermore, we 
argue that a correlated measurement of the difference in the
arrival time of gamma-ray photons and neutrinos emitted from
active galactic nuclei or gamma-ray bursts may provide a signature
of possible violation of Lorentz symmetry. We have found
that this time delay is energy independent, however it has a dependence 
on the chirality of the particles involved. We also briefly discuss the 
known settings where the
mechanism for spontaneous violation of Lorentz symmetry in the context of
string/M-theory may take place.

\vspace*{2.0cm}


\vfill
\eject

\setcounter{page}{2}
\pagestyle{plain}

\setcounter{equation}{0}
\setlength{\baselineskip}{0.8cm}
\section{Introduction}
 
Lorentz invariance is one of the most fundamental symmetries of physics 
and underlies all known physical descriptions of nature. However, more 
recently, there has been evidence, in the context of string/M-theory, 
that this symmetry could, at least in principle, be spontaneously broken.    
This raises immediately the issue of investigating this possibility from the
experimental point of view. Observational information on the violation of 
Lorentz symmetry may, of course, provide essential insights into the nature 
of the fundamental theory of unification and hopefully 
allow establishing relevant bounds on its parameters. 

In this work, we shall argue that astrophysics may play an essential role in
this respect. This comes about as it will soon be possible to make correlated
astrophysical observations involving high-energy radiation and neutrinos.
Indeed, it is remarkable that there exists convincing evidence that 
the observed jets of 
active galactic nuclei (AGN), powered by supermassive black holes at their 
core, are quite efficient cosmic proton accelerators.    
The photoproduction of neutral pions by accelerated protons are assumed to be
the source of the highest-energy photons through which most of the luminosity
of the galaxy is emitted. The decay of charged pions with the ensuing
production of neutrinos is another distinct signature of the proton induced
cascades \cite{Mannheim}. Moreover, there is a consensus that  
estimates of the neutrino flux are fairly model independent and that reliable
upper bounds can be established \cite{Waxman1}. Since gamma-ray bursts (GRB) 
have also
been suggested as a possible source of high-energy neutrinos \cite{Waxman2} 
the mentioned upper bounds are also valid for those sources. To further
deepen the knowledge of these sources and also because the 
cosmic-ray energy spectrum extends to energies higher than $10^{20}~eV$, 
large area ($\sim km^2$) high-energy neutrino telescopes are under 
construction (see e.g. \cite{Gaisser}). These telescopes will allow for 
obtaining information 
that is intrinsically correlated with the gamma radiation emitted by 
AGN and GRB. On the other hand, it has already been pointed out that 
astrophysical observations
of faraway sources of gamma radiation could provide important hints
on the nature of gravity-induced wave dispersion in vacuum \cite{Amelino1,
Biller, Amelino2, Ellis} and hence on physics beyond the standard model (SM). 
In here, we will show that delay measurements 
in the arrival time of correlated sources of gamma radiation 
and high-energy neutrinos can, when considered in the context of 
a Lorentz-violating extension of the SM \cite{Colladay}, 
help in setting a relevant
limit on the violation of that fundamental symmetry.  
We shall further relate our results with the recently discussed 
limit on the violation of Lorentz symmetry from the observations of
high-energy cosmic rays beyond the Greisen-Kuzmin-Zatsepin (GKZ) 
cutoff \cite{Coleman}.

However radical, the idea of dropping the Lorentz symmetry has been 
repeatedly considered in the literature. Indeed, a background or constant
cosmological vector field has been suggested as a  
way to introduce our velocity with respect to a preferred frame of reference 
into the physical description \cite{Phillips}. It has also been
suggested, based on the behaviour of the renormalization group
$\beta$ function of non-Abelian gauge theories,
that Lorentz invariance could be just a low-energy symmetry
\cite{Nielsen}. Furthermore, higher-dimensional theories of gravity 
that are not locally Lorentz invariant have been  
considered in order to obtain light fermions in chiral representations
\cite{Weinberg}.

The  breaking of Lorentz symmetry due to
nontrivial solutions of string field theory was first discussed
in Refs. \cite{Kostelecky1}. These nontrivial solutions
arise in the context of the string field theory of open strings and may
have striking implications at low-energy. For instance, assuming that 
the contribution 
of Lorentz-violating interactions to the vacuum energy is about half of 
the critical density was shown to allow concluding on the existence 
of quite feeble tensor mediated interactions
in the range of about $10^{-4}~m$ \cite{Bertolami1}. 
Furthermore, Lorentz violation 
may be a factor in the breaking of conformal symmetry and this together with 
inflation may lie at the origin of the primordial magnetic fields which are 
required to explain the observed galactic magnetic field \cite{Bertolami2}.
Of course, putative violations of the Lorentz invariance may contribute to
in the breaking of CPT symmetry \cite{Kostelecky}. 
Interestingly, this  
possibility can be verified experimentally in 
neutral-meson \cite{Kostelecky2} 
experiments\footnote{These CPT violating effects are  
unrelated to those that are due to possible nonlinearities in  
quantum mechanics, presumably arising from quantum gravity, which were  
already investigated by the CPLEAR Collaboration \cite{Adler}.}, Penning-trap 
measurements \cite{Bluhm1} and hydrogen-antihydrogen spectroscopy 
\cite{Bluhm2}. 
Moreover, the breaking of CPT symmetry also allows for 
an explanation of the baryon asymmetry of the Universe. 
Tensor-fermion-fermion interactions expected in the low-energy limit of 
string field theories give rise in the early Universe, 
after the breaking of the Lorentz and
CPT symmetries, to a chemical potential that creates in equilibrium a
baryon-antibaryon asymmetry in the presence of baryon number 
violating interactions \cite{Bertolami3}.

Limits on the violation of Lorentz symmetry have been 
directly sought through
laser interferometric versions of the Michelson-Morley experiment,  
where comparison is made between the velocity of light, $c$, and the  
maximum attainable velocity of massive particles, $c_i$, up to  
$\delta \equiv |c^2/c_{i}^2 - 1| < 10^{-9}$ \cite{Brillet}. 
More accurate tests can be performed 
via the so-called Hughes-Drever experiment \cite{Hughes, Drever}. In the 
latter type of measurement, one searches for
the time dependence of the quadrupole splitting of nuclear Zeeman levels
along Earth's orbit and that allows for the achievement of impressive 
limits, for instance,
$\delta < 3 \times 10^{-22}$ \cite{Lamoreaux}. Actually, a more 
recent assessment of these experiments reveals that more stringent 
bounds, up to eight orders 
of magnitude, can be reached \cite{Kostelecky3}.
From the astrophysical side, limits on the violation of momentum 
conservation and the 
existence of a preferred reference frame can also be established from 
bounds on the
parametrized post-Newtonian parameter, $\alpha_{3}$. This parameter vanishes 
identically in general relativity and can be accurately determined 
from the pulse period of pulsars and millisecond  
pulsars \cite{Will, Bell}. The most 
recent limit, $|\alpha_{3}| < 2.2 \times 10^{-20}$ \cite{BellD}, 
indicates that Lorentz symmetry holds up to this level.

In what follows we shall compute the corrections to the dispersion relation
arising from a Lorentz-violating extension of the SM and 
confront it with the evidence on the violation of Lorentz 
invariance as arising from cosmic ray physics. Moreover, we shall show 
that these corrections induce a time delay in the arrival of signals 
carried by different particles from 
faraway sources. We also find that this time delay 
is energy independent, but it has a dependence on the chirality of the 
arriving particles.

\section{Lorentz-Violating Extension of the Standard Model and Dispersion 
Relation}

It is widely believed that the SM, although quite 
successful from the phenomenological viewpoint, is a 
low-energy approximation of a more fundamental theory where unification with 
gravity is achieved and the hierarchy problem solved. It is quite 
conceivable that, in this most likely higher-dimensional underlying theory, 
fundamental symmetries, such as 
CPT and Lorentz invariance, may undergo spontaneous symmetry breaking. 
The fact that within string/M-theory, currently the most promising proposal 
for a fundamental theory, a mechanism where spontaneous breaking of 
Lorentz symmetry is known \cite{Kostelecky1, Kostelecky, Dvali},  
indicates that the violation of those symmetries might actually occur and that 
its implications may be expected.

A priori, there is no reason for this breaking not to extend into the
four-dimensional spacetime. If this is so, CPT and Lorentz
symmetry violations will be likely to occur within the SM and its 
effects might be detected.
In order to account for the CPT and Lorentz-violating effects an extension 
to the minimal SU(3) $\otimes$ SU(2) $\otimes$ U(1) SM has been developed 
\cite{Colladay} based on the idea that CPT and Lorentz-violating terms
might arise from the interaction of tensor fields to Dirac fields 
when Lorentz tensors acquire nonvanishing vacuum expectation 
values. Interactions of this form are expected to arise from the 
string field trilinear self-interaction, as in the 
open string field theory \cite{Kostelecky1, Kostelecky}. 
These interactions may also emerge from the scenario where our world 
is wrapped in a brane and this is allowed to tilt \cite{Dvali}.
Aiming to preserve power-counting renormalizability of the SM, 
only terms involving operators of mass dimension four
or less are considered in this extention. In this 
work, only the fermionic sector of the extension discussed in \cite{Colladay} 
will be considered~\footnote{We shall suppose that the propagation features 
of photons are unaltered and hence that a Lorentz-violating extension of the 
SM gauge sector is unnecessary. Even though this possibility has been 
discussed 
in \cite{Colladay}, the phenomenological restrictions are  
quite severe, 
at least in what concerns the term that gives origin to a cosmological 
birefringence \cite{Carroll}.}. 
This sector includes both leptons and quarks, since
SU(3) symmetry ensures violating extensions to be colour-independent. The
fermionic sector contains CPT-odd and CPT-even contributions to
the extended Lagrangian, which are given by \cite{Colladay}

\beq
{\cal L}^{\rm CPT-odd}_{\rm Fermion} = - a_{\mu} \overline{\psi} 
\gamma^{\mu} \psi -
b_{\mu} \overline{\psi} \gamma_5 \gamma^{\mu} \psi \quad,
\label{1.1}
\eeq

\beq
{\cal L}^{\rm CPT-even}_{\rm Fermion} = \half i c_{\mu\nu} \overline{\psi}
\gamma^{\mu} \rl \psi + \half i d_{\mu\nu} \overline{\psi} \gamma_5
\gamma^{\mu} \rl  \psi - H_{\mu\nu} \overline{\psi} 
\sigma^{\mu\nu} \psi
\quad,
\label{2.1}
\eeq
where the coupling coefficients $a_{\mu}$ and $b_{\mu}$ have dimensions of
mass, $c_{\mu\nu}$ and $d_{\mu\nu}$ are dimensionless and can have both
symmetric and antisymmetric components, and $H_{\mu\nu}$ 
has dimension of mass and is antisymmetric. All the Lorentz-violating 
coefficients are Hermitian.
These parameters are flavour-dependent and some of them 
may induce flavour changing 
neutral currents when nondiagonal in flavour.
  
The Langrangian density of the fermionic sector including
Lorentz-violating terms reads:

\beq
{\cal L} = \half i \overline{\psi}~\gamma_{\mu} \rl \psi - 
a_{\mu} \overline{\psi}~\gamma^{\mu} \psi - b_{\mu} \overline{\psi} 
~\gamma_5 \gamma^{\mu} \psi
+ \half i c_{\mu\nu} \overline{\psi}~\gamma^{\mu} \rln\psi 
+ \half i d_{\mu\nu} \overline{\psi}~\gamma_5 \gamma^{\mu} \rln \psi 
- H_{\mu\nu} \overline{\psi}~\sigma^{\mu\nu} \psi - m \overline{\psi} 
\psi \quad,
\label{2.2}
\eeq
where only kinetic terms are kept, as we are interested in
deducing the free particle energy-momentum relation. 

From the above Lagrangian density, we can get the Dirac-type
equation
\beq
\left[i \gamma^{\mu} (\prt_{\mu} + (c_{\mu}^{~\al} - d_{\mu}^{~\al} 
\gamma_5)~
\prt_{\al}) - a_{\mu} \gamma^{\mu} - b_{\mu} \gamma_5 \gamma^{\mu} - 
H_{\mu\nu} \sigma^{\mu\nu} -m \right] \psi = 0 \quad.
\label{2.3}
\eeq

In order to obtain the corresponding Klein-Gordon equation, we multiply Eq.
(\ref{2.1}) from the left by itself with
an opposite mass sign yielding
\begin{eqnarray}
&&\biggl[[i (\prt_{\mu} + c_{\mu}^{\al} \partial_{\al}) - a_{\mu}]^2 
+(d_{\mu}^{\al} \prt_{\al})^2 - b^2
- m^2 - i \si^{\mu\rh} [i \prt_{\mu} c_{\rho}^{~\beta} \prt_{\beta}
+ i c_{\mu}^{~\alpha} \prt_{\alpha} i \prt_{\rho} 
+ i c_{\mu}^{~\alpha} \prt_{\alpha} i c_{\rho}^{~\beta} \prt_{\beta}
\nonumber \\  
&-& i (\prt_{\mu} + c_{\mu}^{~\alpha} \prt_{\alpha}) i d_{\rho}^{~\beta} 
\ga_5 \prt_{\beta} + i d_{\mu}^{~\alpha} \ga_5 \prt_{\alpha}
i (\prt_{\rho} + c_{\rho}^{~\beta} \prt_{\beta})
- 2 b_{\mu} \ga_5 [i (\prt_{\rh} + c_{\rh}^{\be} \prt_{\be}) - a_{\rh}]]
\nonumber \\    
&-& 2 i (i a_{\mu} \si^{\mu\rh} - b_{\mu} \ga_5 g^{\mu\rh}) 
d_{\rh}^{~\be} \ga_5 \prt_{\be}   
+ \si^{\mu\nu} \si^{\rh\si} H_{\mu\nu} H_{\rh\si} - H_{\rh\si}
(\ga^{\mu} \si^{\rh\si} + \si^{\rh\si} \ga^{\mu}) [i (\prt_{\mu} 
\nonumber \\ 
&+& (c_{\mu}^{~\al} - d_{\mu}^{~\al} \ga_5) \prt_{\al}) - a_{\mu} + b_{\mu}
\ga_5]\biggr] \ps = 0 \quad.
\label{2.4}
\end{eqnarray}

To eliminate the off-diagonal terms, the squaring procedure has to
be repeated. However, as already discussed, since Lorentz symmetry breaking 
effects are quite constrained experimentally (see also 
\cite{Kostelecky, Bertolami1, Colladay} for theoretical discussions), 
violating terms higher than second order will be ignored. After some algebra, 
we find that off-diagonal terms cannot be fully eliminated, but that 
these terms are higher order in the Lorentz-violating parameters. To further
simplify our computation we shall drop $H_{\mu\nu}$. This simplification 
is justifiable as in our phenomenological study we shall only consider the
effect of the timelike components of the Lorentz-violating parameters. Hence,
we obtain, for the Klein-Gordon type equation, up to second 
order in the new parameters:
\begin{eqnarray}
&&\biggl[[(i \prt)^2 + 2i \prt_{\mu} i c^{\mu\al} \prt_{\al} - 2i \prt_{\mu}
a^{\mu} - m^2]^2 
+4 i \prt_{\mu} i d_{\rho}^{~\be} \prt_{\beta} i \prt_{\eta}
i d_{\phi}^{~\delta} \prt_{\delta} (g^{\mu \et} 
g^{\rh\ph} - g^{\mu\rh} g^{\et\ph})
\nonumber \\  
&-& 8 i \prt_{\mu} i d_{\rho}^{~\be} \prt_{\beta} b_{\eta} i \prt_{\phi}
(g^{\mu \rho} g^{\eta\ph} - g^{\mu\phi} g^{\rho\eta})
+ 4 b_{\mu} b_{\et} i \prt_{\rh} i \prt_{\ph} (g^{\mu \et}
g^{\rh\ph} - g^{\mu\rh} g^{\et\ph}) \biggr] \ps = 0 \quad~.
\label{2.5}
\end{eqnarray}

Thus, in the momentum space, we obtain, at the 
lowest nontrivial order, the following relationship:
\begin{eqnarray}
&&(p_{\mu} p^{\mu} + 2 p_{\mu} c^{\mu\al} p_{\al} + 2 p_{\mu} a^{\mu} -
m^2)^2 
+ 4 [p_{\mu} p^{\mu}  d_{\rho}^{~\be} p_{\beta} d^{\rho \delta} p_{\delta} 
- (p_{\mu} d^{\mu \beta} p_{\beta})^2]
\nonumber \\
&+& 8 (p_{\mu} p^{\mu}  d_{\eta}^{~\be} p_{\beta} b^{\eta} -
p_{\mu} d^{\mu\be} p_{\beta} b_{\eta} p^{\eta})
+ 4 [b_{\mu} b^{\mu} p_{\nu} p^{\nu} - (b_{\mu} p^{\mu})^2] = 0 \quad.
\label{2.6}
\end{eqnarray}

Hence, the dispersion relation arising from the 
Lorentz-violating extension of the SM is given by
\begin{eqnarray}
&&p_{\mu} p^{\mu} - m^2 = -2 p_{\mu} c^{\mu\al} p_{\al} - 2 p_{\mu} a^{\mu}  
\pm 2 \biggl[(p_{\mu} d^{\mu \beta} p_{\beta})^2 - p_{\mu} p^{\mu} 
d_{\eta}^{~\be} p_{\beta} d^{\eta \delta} p_{\delta}
\nonumber \\
&+&2 (p_{\mu}  d^{\mu\be} p_{\beta} b_{\rho} p^{\rho} - 
p_{\mu} p^{\mu} d_{\eta}^{~\be} p_{\beta} b^{\eta})
- b_{\mu} b^{\mu} p_{\nu} p^{\nu} + (b_{\mu} p^{\mu})^2
\biggr]^{1/2} \quad,
\label{2.7}
\end{eqnarray}
where the $\pm$ sign refers to the fact that the effects of $b_{\mu}$ and 
$d_{\mu\nu}$ depend on chirality. 

Finally, considering, for simplicity, the scenario where coefficients 
$a_{\mu}$, $b_{\mu}$, $c_{\mu\nu}$, and $d_{\mu\nu}$ have only timelike 
components, it follows that the dispersion relation simplifies to
\beq
p_{\mu} p^{\mu} - m^2 = - 2 c_{00} E^2 - 2 a E \pm 2 (b + d_{00}E) p \quad,
\label{2.8}
\eeq
where we have dropped the component indices of coefficients $a$ and $b$.
From now on we shall also drop coefficient $a$ as it may lead to changing 
flavour neutral currents when more than one flavour is involved. 

In the next section, we shall use the dispersion relation 
Eq. (\ref{2.8}) to see how the
GZK cutoff for ultra-high energy cosmic rays can be relaxed. The 
following discussion is similar to the one described in \cite{Coleman}, 
where it is assumed that the limiting velocities of particles 
in different reference frames are not the same.

\section{Ultra-High Energy Cosmic Rays in the Lorentz-violating Extention
of the Standard Model}

The discovery of the cosmic background radiation has made raising
the question of how the most energetic cosmic-ray particles would be
affected by the interaction with the sea of microwave photons inevitable. 
In fact,
the propagation of the ultra-high energy nucleons is limited
by inelastic impacts with the ubiquitous photons of the background
radiation disabling nucleons with energies above $5 \times 10^{19}~eV$
to reach Earth from further than $50 - 100~Mpc$. This is the well known
GZK cutoff \cite{Greisen}. However, events where the estimated energy of 
the cosmic primaries is beyond the GZK cutoff have been observed 
by different collaborations \cite{Yoshida, Bird, Brooke, Efimov}. 
It has been suggested \cite{Coleman} (see also \cite{Mestres}) 
that slight violations of Lorentz
invariance would cause energy-dependent effects which would suppress processes,
otherwise dynamically inevitable, e.g. the resonant scattering reaction,
\beq
p + \ga_{2.73K} \to \De_{1232} \quad,
\label{3.1}
\eeq
which is at the very core of the GZK cutoff. 
Were this process untenable, the
GZK cutoff would not exist and consequently a cosmological origin for the
high-energy cosmic radiation could be possible \footnote{Actually, it has been
pointed out that the five highest-energy cosmic ray events seem to be closely
correlated in space with cosmologically distant 
compact radio-loud quasars \cite{Farrar}.}. 
As discussed in \cite{Coleman},
this can occur through a change in the dispersion relation for free particles.
We shall see that this is indeed what happens when analyzing process
Eq. (\ref{3.1}) with dispersion relation Eq. (\ref{2.8}).  
Considering a head-on impact of a proton of energy $E$ with a 
cosmic background radiation photon of energy $\om$, the likelihood of the 
process Eq. (\ref{3.1}) would be conditioned by Eq. (\ref{2.8}) to be
\beq
\label{eq}
2\om + E \ge  m_{\Delta} (1 - c^{\Delta}_{00}) \quad.
\label{3.2}
\eeq
Hence, we get the following relationship,
after squaring Eq. (\ref{3.2}) and dropping the $\omega^2$ term:

\beq
2\om + {m_{p}^2 \over 2 E} \ge (c^{p}_{00} - c^{\Delta}_{00}) E 
+ {m_{\De}^2 \over 2 E} \quad,
\label{3.3}
\eeq
which clearly exhibits Lorentz-violating terms.

Let us now compare Eq. (\ref{3.3}) with the results of Ref. \cite{Coleman} and
show that this leads to a bound on $c^{i}_{00}$. To modify the usual 
dispersion relation for free particles, Coleman and Glashow suggest 
assigning a maximal attainable velocity to each particle. 
Therefore, for a given particle $i$ moving freely in the preferred frame, 
which could be thought of as the
one in relation to which the cosmic background radiation is isotropic,
the relevant dispersion relation would be
\beq
E^2 = p^2 c_{i}^2 + m_{i}^2 c_{i}^4 \quad.
\label{3.4}
\eeq 

Hence the likelihood of the process Eq. (\ref{3.1}) to occur
under the conditions described above would depend on 
satisfying the kinematical condition $2\om + E \ge m_{eff}$, 
where the effective mass $m_{eff}$ is given by
\beq
m_{eff}^2 \equiv m_{\Delta}^2 - 
(c_{p}^2 - c_{\De}^2) p^2 \quad,
\label{3.5}
\eeq
the momentum being in respect to the preferred frame.

The likely condition takes then the following form
\beq
2\om + {m_{p}^2 \over 2 E} \ge (c_{p} - c_{\De})E 
+ {m_{\De}^2 \over 2 E} \quad,
\label{3.7}
\eeq
where the term proportional to $c_{p} - c_{\De}$ is clearly 
Lorentz-violating. If the difference in the maximal velocities exceeds 
the critical value 
\beq
\de (\om) = {2 \om^2 \over m_{\De}^2 - m_{p}^2} \quad,
\label{3.8}
\eeq
then reaction Eq. (\ref{3.1}) would be forbidden and consequently the 
GZK cutoff
relaxed. For photons of the microwave background, $T = 2.73~K$, and 
$\omega_{0} \equiv kT = 2.35 \times 10^{-4}~eV$, this condition would be
\beq
c_{p} - c_{\De} = \de (\om_{0}) \simeq 1.7 \times 10^{-25} \quad,
\label{3.9}
\eeq
which is quite a striking limit on the violation of the Lorentz symmetry, 
even though it is valid only for the process in question. Similar bounds for 
other particle pairs, although less stringent, were discussed in 
\cite{Coleman, Halprin}. 

Finally, thecomparison of Eq. (\ref{3.7}) with Eq. (\ref{3.3}) 
gives for $\Delta c_{00}$:
\beq
c^{p}_{00} - c^{\Delta}_{00} \simeq 1.7 \times 10^{-25}  \quad.
\label{3.10}
\eeq

Thus, we see that the Lorentz-violating extension of the SM can explain 
the violation of the GZK cutoff and account for the 
phenomenology of ultra-high energy cosmic rays via the bound on 
$\Delta c_{00}$ given by Eq. (\ref{3.10}). Of course, the situation would 
be more 
complex if the Lorentz-violating parameters were allowed to have spacelike 
components which would lead to direction and helicity-dependent effects.

\section{An astrophysical test of Lorentz Invariance}

Let us turn to the discussion of a possible astrophysical test of Lorentz
invariance. From eq. (\ref{3.10}) we see that
$\Delta c_{00} \simeq \epsilon$, where $\epsilon$ is a small constant 
specific of the process involved (cf. eq. (\ref{3.10})) 
and, for instance, $\vert \epsilon \vert \lsim few \times 10^{-22}$ 
from the search of neutrino oscillations \cite{Brucker, Glashow}. 
In what concerns signals simultaneously emitted by faraway sources, 
the resulting effect in the propagation velocity of 
particles with energy, $E$, and momentum, $p$, is given 
in the limit where $m << p, E$ or for massless 
particles by $c_i = c [1 - (c_{00} \pm d_{00})_{i}]$. 
Hence, for sources at a distance $D$, the relative delay in the arrival 
time will be given by
\beq
\Delta t \simeq {D \over c} [(c_{00} \pm d_{00})_{i} - 
(c_{00} \pm d_{00})_{j}] \equiv \epsilon_{ij}^{\pm} {D \over c} \quad,
\label{4.1}
\eeq
where we have defined a new constant, $\epsilon_{i j}^{\pm}$, 
involving a pair of particles. This time delay may, despite being given by the 
difference between two quite small numbers, be measurable for sufficiently 
far away sources. Moreover, our result indicates that the estimated 
time delay is energy independent, in 
opposition to what one could expect from general arguments \cite{Amelino1,
Amelino2, Ellis}. We have also found that the time delay has an
interesting dependence on the chirality of the particles involved. 
In the next section, we shall discuss how to estimate the 
scales involved in the observational value of $c^{i}_{00}$ 
(and $d^{i}_{00}$ if $d^{i}_{00} \sim c^{i}_{00}$).   

Therefore if, for instance, the signals from faraway sources were, 
as suggested in the introduction, from an AGN $TeV$ gamma-ray flare 
and the genetically 
related neutrino emission, then we should expect for the time delay, 
if as justified above that the photon propagation is unaltered,  
\beq
\Delta t \simeq (c_{00} \pm d_{00})_{\nu} {D \over c} \quad,
\label{4.2}
\eeq
assuming that the neutrinos are massless, an issue that will hopefully be 
settled experimentally in the near future. It is worth stressing 
that even before that, the effect of neutrino masses and other 
intrinsic effects related to the nature of the neutrino 
emission processes can, 
at least in principle, be extracted from the 
data of several correlated detections of $TeV$ gamma-ray 
flares and neutrinos if a systematic delay of 
neutrinos is observed. However, the main point here is that a 
nonvanishing time delay can be regarded, up to neutrino
mass effects and neutrino emission processes, as direct evidence for  
a violation of Lorentz symmetry and, as already pointed out in 
the introduction, neutrino telescopes
will soon be available for the investigation of correlated detections.
Furthermore, the available knowledge of AGN phenomena and our
confidence in the astrophysical methods available to determine their 
distance from us make it reasonable
to believe that the time delay strategy may realistically 
provide relevant limits on the
violation of Lorentz symmetry. Of course, the same arguments may very well 
apply to GRB; however, the lack of
a deeper understanding of these transient phenomena introduces further 
unnecessary uncertainty, even though many properties of their sources can be 
understood from the observation of their afterglows. It is also important to
point out that limits involving photons and neutrinos are currently unknown and
that a difference between neutrinos and antineutrinos is expected if 
$d^{i}_{00}$ is nonvanishing.
We could also add that, based on the analysis of Ref. 
\cite{Kostelecky3} involving the full set of Lorentz-violating parameters, 
we expect our results to remain unaltered, at least at high energies, 
if the parameter $H_{\mu \nu}$ were to be held in our calculations.

\section{Discussion and Conclusions}
 
In summary, we have shown that parameters of the Lorentz-violating
extension of the SM proposed in Ref. \cite {Colladay} can be related to the
phenomenology of ultra-high energy cosmic rays with the conclusion that,
as in \cite{Coleman}, it may lead to the suppression of processes responsible
for the GZK cutoff. This is a crucial argument for an extragalactic 
origin of high-energy cosmic rays. Furthermore, we have found that 
the relevant Lorentz-violating parameter is, at high energies, $c^{i}_{00}$ 
so that $\Delta c_{00} \simeq \epsilon$ with 
$\vert \epsilon \vert \lsim few \times 10^{-22}$ from neutrino physics 
and $\vert \epsilon \vert \lsim 10^{-25}$ from the ultra high-energy
cosmic-ray physics.
Actually, it is possible to estimate the typical scales involved in the 
problem assuming that the 
source of Lorentz symmetry violation is due to nontrivial solutions 
in string field theory. Indeed, these
solutions imply that Lorentz tensors acquire vacuum expection values as  
Lorentz symmetry is spontaneously broken due to string-induced 
interactions \cite{Kostelecky1, Kostelecky}. 
A sensible parametrization for these expectation values and hence for 
$c^{i}_{00}$ 
would be the following for a fixed energy scale, E,
\cite{Kostelecky, Bertolami3}:
\beq
c^{i}_{00} \simeq {\vev{T} \over M_{S}} = 
\lambda_{i} \left({m_{L} \over M_{S}}\right)^{l} 
\left({E \over M_{S}}\right)^{k}  \quad,
\label{6.1}
\eeq
where $T$ denotes a generic Lorentz tensor, $\lambda_{i}$ is presumably
an order one flavour-dependent 
constant\footnote{Another scenario would be for $\lambda_{i}$
to be of the order of the respective Yukawa coupling 
\cite{Kostelecky}.}, $m_{L}$ is a light 
mass scale, $M_{S}$ is a string scale presumably close to Planck's mass 
or a few orders 
of magnitude below it, and $k, l$ are 
integers indicating the order of the string corrections to low-energy 
physics. Thus, in the
lowest non-trivial order, $k = 0,~l = 1$ ($k=l=0$ is already 
excluded experimentally), $c^{i}_{00} = \lambda_{i}
\left({m_{L} \over M_{S}}\right)$ and for different $\lambda_{i}$ constants
$\epsilon \simeq \left({m_{L} \over M_{S}}\right)$.
This result corresponds, in its essential lines, to the one we have obtained
from working out the implications of the Lorentz-violating
extension of the SM in the context of ultra-high energy cosmic-ray 
phenomenology. Furthermore if, for instance, $\epsilon \lsim 10^{-23}$,
then it follows that $m_{L} \sim 10^{2}~KeV$ for $M_{S} \simeq M_P$ or 
$m_{L} \sim 10^{2}~eV$ if $M_{S} \simeq few \times 10^{16}~GeV$ \cite{Witten}.
Estimates for $m_{L}$ would clearly change by many orders of magnitude if 
$\lambda_{i}$ were of the order of the Yukawa coupling. In either case, our 
main conclusion is that choice  $k =0,~l = 1$ implies that the time delay 
in the arrival 
of signals from faraway sources is, as discussed above, energy independent.
We have found, however, an interesting dependence on the chirality of 
particles involved. 

Naturally, another scenario would emerge from a different choice of integers
$k, l$. For instance, the choice $k=2$ and $l=0$, the relevant choice in the 
CPT symmetry violating baryogenesis
scenario \cite{Bertolami3}, where the energy should, in this case, 
be related to
the early Universe temperature. This would imply that the time delay in the 
arrival of signals from faraway sources 
would be proportional to the square of the energy. This choice would also 
lead to the
conclusion that Lorentz-violating effects, whether due to string physics 
or quantum gravity, are quadratic in the energy. Interestingly,
similar conclusions concerning the order of quantum gravity low-energy 
effects are reached from the study of corrections to the 
Schr\"odinger equation arising from quantum gravity in the mini-superspace
approximation \cite{Bertolami4}.    
   
Another setting allowing for the spontaneous breaking of the Lorentz symmetry
is the so-called braneworld \cite{Dvali}. In this scenario, 
SM particles lie on a 
three-brane, $\phi(x)$, embedded in spacetime, with possibly 
large compact extra dimensions, 
whereas gravity propagates in the bulk. Thus, a tilted brane 
induces rotational 
and Lorentz noninvariant terms in the four-dimensional effective theory
as brane-Goldstones couple to all particles on the brane via an induced metric
on the brane. This will lead to operators of the form
\beq
\prt_{\mu} \phi~\prt_{\nu} \phi~\overline{\psi}~ 
\gamma^{\mu}  \prt^{\nu} \psi + \prt^{\mu} \phi~\prt_{\nu} \phi~
F_{\mu \alpha} F^{\nu \alpha}... \quad,
\label{6.2}
\eeq  
which closely resemble the Lorentz-violating terms in the SM extention. 
As before, phenomenology sets tight constraints on this scenario, which is 
however, currently unable to establish whether the 
breaking of Lorentz invariance, if
observed at all, has its origin in the nonperturbative nature of branes 
or if it arises from the perturbative string field theory 
scenario described above.  
The former alternative could possibly be associated with a $M_S$ scale that 
is a few orders of magnitude below the Planck scale, while the latter with a
$M_S$ that should be associated with the Planck scale itself.

Finally, we could say that based on our results, we have outlined a strategy 
to establish to what extent Lorentz invariance is violated
from the observation of the 
time delay in the detection of $TeV$ gamma-ray flares and neutrinos from
AGN. Our analysis reveals that the time delay has a dependence on the 
chirality of the particles involved, but is energy independent, 
contrary to what one could expect from general arguments. In either case, 
if ever observed, a time delay in the arrival of signals from 
faraway sources would be, up to neutrino mass effects and other features
related to the nature of the neutrino emission, strong 
evidence of quite new physics beyond the SM.


\vspace*{2.0cm}

{\large\bf Acknowledgments} 

\noindent
One of us (O.B.) would like to thank the hospitality of 
Department of Physics of the New York University where part of this work was 
carried out and Funda\c c\~ao para a Ci\^encia e a Tecnologia (Portugal) 
for the financial support under grant No. BSAB/95/99. It is a pleasure to 
thank Alan Kosteleck\'y for important comments and suggestions. We also 
thank A. Halprin for comments.  

\newpage


\end{document}